\documentclass[12pt]{article}

\usepackage{latexsym}

\font\tenmsbm=msbm10 scaled 1200
\font\sevenmsbm=msbm9
\newfam\msbmfam
\textfont\msbmfam=\tenmsbm
\scriptfont\msbmfam=\sevenmsbm

\makeatletter
\@addtoreset{equation}{section}
\makeatother

\newcommand{\eref}[1]{(\ref{#1})}

\newcommand{\beq}{\begin{equation} }
\newcommand{\eeq}{\ensuremath {\end{equation}} }
\newcommand{\bea}{\begin{eqnarray} }
\newcommand{\eea}{\end{eqnarray} }
\newcommand{\pa}{\partial}
\newcommand{\ve}{\varepsilon}
\newcommand{\muh}{\ensuremath{\overline{\mu}} }
\newcommand{\nuh}{\ensuremath{\overline{\nu} }}

\newcommand{\yh}{\ensuremath{\hat{y} }}

\newcommand{\uh}{\ensuremath{\hat{u} }}
\newcommand{\gbar}{\ensuremath {\overline{g}} }
\newcommand{\Gbar}{\ensuremath {\overline{G}} }
\newcommand{\hb}{\ensuremath {\overline{h}} }

\newcommand{\xten}{x^{10}}

\begin{document}

\begin{titlepage}

\begin{flushright}
Preprint DFPD 01/TH/37\\
March 2002\\
\end{flushright}

\vspace{0.5truecm}

\begin{center}

{\Large \bf $NS\, 5$--branes in IIA supergravity and gravitational
anomalies}

\vspace{2cm}
$ \mbox{M. Cariglia}^{* \dag ,}$\footnote{mc355@cam.ac.uk}  
$\mbox{ and K. Lechner}^{\dag ,}$\footnote{kurt.lechner@pd.infn.it} 

\vspace{2cm}

  $^* ${\it DAMTP, Centre for Mathematical Sciences, Cambridge University 
      \smallskip 
      
       Wilberforce Road, Cambridge CB3 OWA, UK }

\vspace{1cm}

$^{\dag}${\it Dipartimento di Fisica, Universit\`a degli Studi di Padova,
    \smallskip

    and

    \smallskip

    Istituto Nazionale di Fisica Nucleare, Sezione di Padova,

    Via F. Marzolo, 8, 35131 Padova, Italia
   }

\vspace{2cm}

\begin{abstract}
\vskip0.2truecm
We construct a gravitational--anomaly--free effective action for the
coupled system of IIA $D=10$ {\it dynamical} supergravity
interacting with an $NS5$--brane. The $NS5$--brane is considered as
{\it elementary}  in that the associated current is a
$\delta$--function supported on its worldvolume.
Our approach is based on a Chern--kernel which encodes  
the singularities of the three--form field strength  
near the brane in an $SO(4)$--invariant way and provides a solution
for its Bianchi identity in terms of a two--form potential.
A dimensional reduction of the recently constructed anomaly--free effective
action for an elementary $M5$--brane in $D=11$ is seen to reproduce  
our ten--dimensional action. The 
Chern--kernel approach provides in particular a concrete realization of
the anomaly cancellation mechanism envisaged by Witten.

\vspace{0.5cm}

\end{abstract}

\end{center}
5--branes, Chern--kernels, anomalies.
\end{titlepage}

\newpage

\baselineskip 6 mm

\section{Introduction}

It is a well known fact that IIA string theory admits 1--branes
and 5--branes as elementary excitations, and that its low energy
limit is given by IIA supergravity. The bosonic sector of IIA
supergravity is given by the metric $g_{\mu\nu}$, the dilaton
$\Phi$, a one--form $A_1$, a two--form $A_2$ and a three--form $A_3$,
while the bosonic fields on the $NS$--5 brane are the coordinates
$x^\mu (\sigma )$, a scalar $a_0(\sigma )$ and a chiral two--form
$a_2(\sigma)$. The chiral fields on the 5--brane lead to
gravitational anomalies whose polynomial has been calculated in
\cite{Witten:fivebrane}; it is given by
 \beq \label{anomaly}
2\pi\left( X_8 +\frac{1}{24}\,\chi_4\chi_4\right),
 \eeq where
$X_8={1\over 192 (2\pi)^4}(tr R^4-{1\over 4}(tr R^2)^2)$ is the
$SO(1,9)$ target space anomaly polynomial and $\chi_4$ is the
Euler characteristic of the $SO(4)$--normal bundle of the brane.
We will use the following notation for descent equations: $\chi_4 =
d\chi_3$, $\delta\chi_3 =d\chi_2$, and similarly for all other
polynomials.

While the target space anomaly can be cancelled through a standard
Green--Schwarz mechanism the normal anomaly requires an inflow
mechanism, relying on the basic equation
\begin{equation}
 dH_3 = g J_4,
\label{eq:dH_3}
\end{equation}
where $H_3$ is the curvature associated to $A_2$, $g$ the charge
of the brane and the four--form $J_4$ is the current associated to
the {\it elementary} brane, i.e. a Dirac $\delta$--function with
support on the brane. For $g=0$ one recovers pure
ten--dimensional bosonic IIA supergravity and eq. \eref{eq:dH_3}
allows to introduce a potential $A_2$ according to $H_3 =dA_2$. When
$g \neq 0$ instead, the introduction of a potential becomes problematic
because the r.h.s. of \eref{eq:dH_3} is different from zero.
On the other hand, a consistent definition of a potential for
$H_3$ becomes of fundamental importance if one wants to write an
action and, consequently, if one considers the issue of anomaly
cancellation by some sort of inflow mechanism.

Usually the system IIA supergravity +  $NS\, 5$--brane has only
been considered in the framework of the $\sigma $--model, where
target space fields act as sources for the fields living on the
brane (direct coupling), but there is no influence of the brane on
target space fields (back coupling): they satisfy the equations of
motion of pure supergravity. It is clear that the modified Bianchi
identity \eref{eq:dH_3} can not be the unique back coupling
equation, because it induces also modifications to the Bianchi
identities and equations of motion for the other bosonic
supergravity fields; principal aim of the present paper is to
present these modified equations and to write down the corresponding
action in compatibility with anomaly cancellation. This requires
in particular to solve the problem of a potential $A_2$ for $g\neq 0$.

The issue of anomaly cancellation for $IIA$ $NS5$--branes has been
addressed in two related ways up to now. The author of
\cite{Witten:fivebrane} considers an ``almost $\delta$-like"
source by arguing that the $NS5$--brane $\delta$--like current
$J_4$ should be replaced by a {\it cohomologically equivalent} closed
form $\widetilde J_4$ with
support in a small neighborhood of the brane. In order to perform
the pullback on the 5--brane worldvolume $M_6$ of
equation \eref{eq:dH_3} he exploits the property that there exists a
regular cohomological representative of $J_4$ such that its
pullback is $\chi_4$,
\beq
\label{wi} \widetilde J_4^{(0)}=\chi_4,
\eeq
where with the subscript $(0)$ we indicate the pullback of a
form on $M_6$. Then he argues that $\chi_4$ should be considered as the
``finite part" of the $\delta $-function in $J_4$, and that
\eref{wi} ``should be taken as part of the definition of the
brane". This leads in particular to the pullback equation
\beq
\label{pullback} dH_3^{(0)}=g\chi_4.
\eeq
He proposes then the
following counterterm in the action to cancel the normal bundle
anomaly
\beq \label{counter}
 \frac{\pi}{12g} \int_{M_6} H_3^{(0)} \chi_3.
\eeq

Unfortunately for an {\it elementary} brane, with a $\delta$--like support, 
this
counterterm is ill--defined because then $J_4$ as well as $H_3$ do not
admit pullbacks on $M_6$. In particular, for its variation under $SO(4)$
one would {\it formally} obtain 
$$
\delta\int_{M_6} H_3^{(0)} \chi_3=-\int_{M_6}dH_3^{(0)}\chi_2
=-g\int_{M_6}J_4^{(0)}\chi_2=-g\int_{M_{10}}J_4J_4\chi_2,
$$
but, since locally we have $J_4=d^4u\,\delta^4(u)$ where the $u^r$ are
normal coordinates, 
the product $J_4J_4$ is 
meaningless, as is the above counterterm. To obtain the desired anomaly 
cancelling term one has to enforce, once more, a 
{\it cohomological} relation: $J_4J_4 \sim J_4\chi_4$ \cite{BottTu,CheungYin}.
Notice however that in the framework of
\cite{Witten:fivebrane} using \eref{pullback} one can introduce a pullback
potential according to $H_3^{(0)}=dA_2^{(0)}+g\chi_3$ and rewrite
\eref{counter} in the form
$$
\frac{\pi}{12g} \int_{M_6} A_2^{(0)}\, \chi_4,
$$
which cancels still the normal bundle anomaly if one enforces the
transformation law
\beq
\label{trans}
\delta A_2^{(0)}=-g\chi_2.
\eeq
This second form of the counterterm may be well defined even for an
elementary brane, but in this case one has to face the problem of how to
introduce a {\it target space} form $A_2$ which gives rise to \eref{trans}
in a consistent way; in particular, for continuity reasons, one must now 
have also
a non trivial target space transformation law: $\delta A_2\neq 0$.
This is one of the main problems solved in the present paper.

The aim of the paper \cite{BeckerAndBecker} instead was a derivation of
the counterterm \eref{counter} from the eleven--dimensional $M5$--brane
effective action proposed in \cite{HarveyMinasianMoore}. Since the
eleven--dimensional 5--brane current $\widetilde J_5$ used in this paper
was a smooth one, the dimensional reduction led again to a smooth current
$\widetilde J_4$ and, once more, to \eref{counter}.
 
The main reason why we insist on a $\delta$--like current
stems from electric--magnetic duality: if one wants that 1--branes
and 5--branes can consistently coexist one has to impose not only
a Dirac quantization condition for their charges but also to
demand that both their currents carry a $\delta$--like support;
this is explained, for example, in \cite{LechnerMarchetti}.
Moreover, solitonic solutions of eleven-- or ten--dimensional
supergravity lead to $\delta$--like supports even for solitonic
5--branes \cite{Gueven}. The requirement of existence of an
effective action for an elementary 5--brane is also in line with
the fact that the derivation of its gravitational anomaly
\eref{anomaly} can be based on a local $\sigma$--model action
where, by definition, the current is strictly a $\delta$--function
\cite{LechnerTonin}.

With these motivations in mind we present in this paper an action
describing the interaction between $IIA$ supergravity and an
$NS5$--brane, which copes consistently with equation
\eref{eq:dH_3} where $J_4$ is considered as a $\delta$--function
on the 5--brane worldvolume, see below for a precise definition.
As anticipated above the first problem one has to solve is the
introduction of a two--form potential $A_2$. Since $J_4$ is a
closed form the first step demands to find a three--form $K_3$
such that
$$
\label{kern}
dK_3=J_4,
$$
because this would then allow to define $A_2$ according to
$$
H_3=dA_2+gK_3.
$$
Since we want $A_2$ to be a regular field near the 5--brane, i.e.
the pullback $A_2^{(0)}$ has to be well defined, the singularities
which are necessarily present in $H_3$ have to be carried entirely
by $K_3$. We will see that this implies that $K_3$ decomposes into
a sum
 $$
 K_3=\chi_3+\omega_3,
 $$
where $\chi_3$ represents a target--space form whose pullback on
$M_6$ is {\it regular} and coincides with the Chern--Simons form
associated to the Euler form. On the other hand, the three--form
$\omega_3$ -- a Chern--kernel -- exhibits a {\it singular and
invariant} behaviour near the 5--brane. Our construction of the
effective action will be based on this decomposition of $K_3$.

As a check of our result we perform a dimensional reduction of the
effective action for an elementary $\delta$--like $M5$--brane in
$d=11$, which has been recently constructed in
\cite{LechnerMarchettiTonin}, and which is based on a
Chern--kernel, too. This action cancels the normal $SO(5)$--bundle
anomaly and relies on an {\it invariant} three--form potential
$A_3$, while the ten--dimensional two--form $A_2$ carries an
anomalous transformation law; so the anomaly cancellation
mechanisms look rather different in ten and eleven dimensions. An
interesting aspect of the reduction regards the way in which the
eleven--dimensional mechanism gives rise to the ten--dimensional
one. Eventually the reduced action agrees with the
ten--dimensional one, we have constructed independently.

An alternative approach for the solution of \eref{eq:dH_3} is
based on Dirac--branes, i.e. unphysical surfaces whose boundary is
the 5--brane. Although being a standard approach, in the present
case it is appropriate for what concerns the equations of motion
but it fails at the level of the action, due the presence of the
anomaly cancelling counterterm; we explain the reason for this
failure in section 3. In some sense the Chern--kernel approach
represents a refinement of the Dirac--brane one.

In order to consider the 5--brane as a $\delta$-like source one
has necessarily to work with distribution valued $p$--forms, so
called ``$p$--currents" \cite{deRham, HarveyLawson}. Differentials of
forms will always be differentials in the sense of distributions
\footnote{In our notation, when Leibnitz' rule is valid, the differential
will act from the right.}. We will always suppose to deal with topologically
trivial spaces, unless otherwise stated, so a  closed $p$--form is
the differential of a $(p-1)$--form.

The rest of the paper is organized as follows: in section
\ref{section:11D} we give a brief selfcontained account of the
eleven--dimensional effective $M5$--brane action, in section
\ref{section:10D} we construct the effective action for IIA
dynamical bosonic supergravity interacting with a bosonic
$NS5$--brane and eventually, in section
\ref{sec:dimensional_reduction}, we consider the issue of
dimensional reduction of the eleven--dimensional theory. Section
5 is devoted to concluding remarks.

\section{The $M5$--brane in eleven dimensions \label{section:11D} }

We present here the basic ingredients of the effective action
for an elementary $M5$--brane, constructed recently in
\cite{LechnerMarchettiTonin}; for more details we refer the reader
to this reference.

\subsection{Poincar\`e--duality and push--forward}

In the presence of a brane one can define differential forms which
live on the brane and differential forms which live on the target
space. The pullback operation associates an $n$--form on the brane
worldvolume to an $n$--form on the target space; the push--forward
operation, instead, associates a form (properly speaking, a
current) on the target space to a form on the worldvolume. Since
we will use this operation frequently we give here its definition
and state its basic properties \cite{deRham,morgan}. We begin by
recalling the definition of  the Poincar\`e--dual of a
$(D-p)$--manifold, open or closed, with worldvolume $M_{D-p}$: it
is the $p$--current $J_p$ --  the $\delta$--function on $M_{D-p}$
-- defined through
$$
\int_{R^D} \Phi_{D-p}J_p=\int_{M_{D-p}} \Phi_{D-p}^{(0)}
$$
for every smooth target space $(D-p)$--form, where
$\Phi_{D-p}^{(0)}$ indicates its pullback.

Given an $n$--form $h_n$ on $M_{D-p}$ instead, we can define its
push--forward to an $(n+p)$--form on target space, which we
{\it indicate} with $``h_nJ_p"$. It is defined through
 \beq
 \label{push}
\int_{R^D} \Phi_{D-p-n}(h_nJ_p)=\int_{M_{D-p}} \Phi_{D-p-n}^{(0)}h_n,
 \eeq
again for every target space form $\Phi_{D-p-n}$. Notice, however, that
the product notation $h_nJ_p$ is formal because this target space form
can not really be factorized. An explicit component expression, following
from \eref{push}, is indeed
\bea
 h_n J_p &=& {1\over (n+p)!(D-n-p)!}\,
 dx^{\mu_1} \cdots dx^{\mu_{n+p}}
\,\varepsilon_{\mu_1\cdots \mu_{n+p}\nu_1\cdots\nu_{D-n-p}}\nonumber\\
& &\cdot \int_{M_{D-p}} E^{\nu_1}\cdots
E^{\nu_{D-n-p}}\,h_n\,\delta^D(x-x(\sigma)),\nonumber
 \eea
where $E^\nu=dx^\nu(\sigma)$, and $x^\nu(\sigma)$ parametrizes the
worldvolume $M_{D-p}$. For $n=0$, $h_n=1$ one obtains a component
expression for $J_p$. The product notation is nevertheless
convenient because the definition \eref{push} implies that
Leibnitz's rule for differentiation holds true,
$d(h_nJ_p)=h_ndJ_p+(-)^pdh_nJ_p$. This is the basic property of
the push--forward that we will use throughout this paper.
In practice the definition of the push--forward implies that the
``product" between the Poincar\`e--dual of a brane, $J_p$, and a
form $h_n$ on that brane yields a well defined target space
current, while this is clearly not true for products of generic
target--space--forms and worldvolume--forms. Another property that
we will use frequently is that
$$
\Phi J_p=\Phi^{(0)} J_p,
$$
whenever the target space form $\Phi$ admits pullback on $M_{D-p}$.
In the rest of the paper these definitions and properties are always
understood.

\subsection{Bianchi identities and Chern--kernels \label{subsec:C-k} }

The bosonic fields of $d=11$ supergravity are the metric
$g_{\muh\nuh}(x)$, ($\muh=0,\cdots,10$) and a three--form
potential $B_3$; the $M5$--brane fields are the coordinates
$x^{\muh}(\sigma)$ and the chiral two--form $b_2(\sigma)$. We set
a bar on eleven--dimensional quantities, to distinguish them from
their ten--dimensional counterparts of next sections.
 
The fundamental equation that
describes the back coupling of the brane to supergravity is
\begin{equation}
 dH_4 = \overline{g} J_5,
\label{eq:dH_4bis}
\end{equation}
where $J_5$ is the $\delta$--function on the 5--brane worldvolume
$M_6$, i.e. its Poincar\`e--dual, and
$\gbar $ is the 5--brane's charge.
This equation should be regarded as the Bianchi identity for
the curvature $H_4$ whose solution amounts to the introduction of a
potential $B_3$.

Introducing a set of normal coordinates
$x^{\muh}\leftrightarrow(\sigma ^i ,y^a )$, ($i=0,\dots ,5$,
$a=1,\dots ,5$) where the $\sigma^i $ are local coordinates on the
brane and the $y^a$ parametrize its normal $SO(5)$--bundle, the
current can be rewritten locally as
\begin{equation}
    J_5 = \frac{1}{5!} \ve^{a_1...a_5}dy^{a_1}...dy^{a_5}
    \delta^5(y).
    \label{eq:J_5_expression}
\end{equation}
On the brane there lives an $SO(5)$--connection $A^{ab}(\sigma )$
that is obtained from the Riemannian connection by embedding the
brane in the target--space. One introduces an arbitrary target--space
extension $A^{ab}(\sigma,y)$ of this connection  subject to the boundary
conditions
 \begin{equation}
    \begin{array}{lr}
    A^{ab}(\sigma ,0)=A^{ab}(\sigma ), &
    \lim_{|y|\rightarrow\infty }A^{ab}(\sigma ,y)=0.
    \end{array}
    \label{eq:connection_extension}
 \end{equation}
This extension ensures the correct fall--off at infinity
of the eleven--dimensional Chern--kernel,
which is the target--space $SO(5)$--invariant four--form
 \begin{equation}
K_4 = \frac{1}{16(2\pi )^2}\,\ve^{a_1...a_5}\,\yh^{a_1}K^{a_2
a_3}K^{a_4 a_5} ,
    \label{eq:K_4a}
 \end{equation}
with
\begin{equation}
    \begin{array}{lcr}
    \yh^a = y^a/\sqrt{y^2}, & K^{ab} = F^{ab} + D\yh^a D\yh^b , &
    D\yh^a = d\yh^a +\yh^b A^{ba} ,
    \end{array}
    \label{eq:K_4b}
\end{equation}
and $F=dA+AA$ is the extended $SO(5)$--curvature.
$K_4$ has two important properties: the first is that as a
distribution it satisfies
\begin{equation}
    \begin{array}{lcr}
    dK_4 =J_5 &\rightarrow & H_4 = dB_3 +\gbar K_4 ,
    \end{array}
    \label{eq:dK_4}
\end{equation}
allowing the introduction of a potential $B_3$ that is {\it
regular on the brane}, and the second is that its (singular)
behaviour near the 5--brane is universal, see below. Since
$B_3^{(0)}$ is a well defined field the curvature of the chiral
two--form potential on the brane can be defined in a standard way
as
 \begin{equation}
    \hb_3 =db_2 + B_3^{(0)},
 \end{equation}
in compatibility with the gauge transformations
\begin{equation}
    \left\{ \begin{array}{l}
            \delta B_3 = d\Phi_2 \\
            \delta b_2 = -\Phi_2^{(0)}.
            \end{array}\right.
       \label{eq:11d_gauge_transf}
\end{equation}

A central role in the construction of the action is played by the
identity
 \bea
 K_4 K_4 &=&\frac{1}{4}\,d\,f_7,\\
 f_7 &\equiv& P_7 + Y_7,
 \label{eq:K_4^2}
 \eea
where $P_7$ is the Chern--Simons form associated to the second
Pontrjagin form of the normal $SO(5)$--bundle,
 $$
P_8={1\over 8(2\pi)^4}\left((tr F^2)^2-2trF^4\right)=
 dP_7,
 $$
and $Y_7$ is an $SO(5)$--invariant form
(see \cite{LechnerMarchettiTonin}). Notice that
due to the distributional nature of $K_4$ Leibnitz' rule does not
hold: the eight--form $K_4K_4$ is closed even if $dK_4\neq0$.

\subsection{Effective action \label{subsec:effective_action_11D} }

The effective action for the system dynamical supergravity +
$M5$--brane is the sum of a local classical part and of a quantum
part:
 \begin{equation}
    \Gamma = \frac{1}{\Gbar}\left( S_{kin} + S_{wz} \right)
    +\Gamma_q.
    \label{eq:11d_action}
 \end{equation}
The quantum contribution $\Gamma_q$ carries the gravitational
anomaly associated to the polynomial \cite{Witten:fivebrane},
 $$
2\pi\left(X_8+\frac{1}{24}P_8\right),
 $$
where $X_8$ is the $SO(1,10)$ target--space anomaly polynomial,
formally identical to the ten--dimensional one given in the
introduction, and $P_8$ is defined above. $\Gbar$ is the
eleven--dimensional Newton's
constant, related to $\gbar$ by \footnote{In our notations
the 5--brane tension is given $T_5={\gbar\over\Gbar}$.}
 \begin{equation}
 \label{quantum}
    2\pi\Gbar = \gbar^3.
 \end{equation}
$S_{kin}$ collects the kinetic terms for the spacetime metric, for $B_3$,
for the coordinates $x^{\muh}(\sigma)$ and for $b_2(\sigma)$, the latter being
written in the PST--approach \cite{PST},
 \begin{equation}
 \label{kin}
S_{kin} = \int_{M_{11}} d^{11} x \sqrt{g}\,R -\frac{1}{2}\int_{M_{11} }\,
H_4 * H_4 \, -\gbar \int_{M_6} d^6 \sigma \sqrt{g_6}\left( {\cal
L}(\widetilde{h} )+\frac{1}{4}
    \widetilde{h}^{ij}h_{ij} \right).
 \end{equation}
The kinetic terms for $b_2$ produce a generalized self--duality
condition for $\hb_3$ which involves the Born-Infeld lagrangian
 \begin{equation}
    {\cal L}(\widetilde{h})=\sqrt{det\left( \delta_i^{\, j}
    +i\widetilde{h}_i^{\, j}\right) },
    \label{BIL}
 \end{equation}
with $h_{ij}=v^k \hb_{ijk}$, $\widetilde{h}_{ij}=v^k \left(
* \hb \right)_{ijk}$, $v_k = \pa_k \alpha/\sqrt{-(\pa \alpha)^2 }$,
and $\alpha(\sigma )$ is a non propagating scalar auxiliary field.
$g_6$ is minus the determinant of the induced metric on the 5--brane.

The Wess--Zumino term is written as the integral of an
eleven--form $S_{wz}=\int_{M_{11} } L_{11}$, with
 \begin{eqnarray}
  L_{11}&=&\frac{1}{6}B_3dB_3dB_3 -\frac{\gbar}{2}\,d b_2 B_3^{(0)}J_5
+\frac{\gbar}{2}\,B_3dB_3K_4 + \nonumber \\
&+& \frac{\gbar^2}{2}\,B_3K_4K_4 + \frac{\gbar^3}{24}K_4 f_7 +
   {2\pi\Gbar\over\gbar}X_7 H_4 \label{eq:L_11}.
\end{eqnarray}
Since under an $SO(5)$--transformation we have $\delta f_7=dP_6$,
$\delta X_7=dX_6$
it is immediately checked that ${1\over\Gbar}S_{wz}$ cancels the
gravitational anomaly carried by $\Gamma_q$, thanks to \eref{quantum}.

The classical action $S_{kin}+S_{wz}$ allows to derive the classical
equations of motion for $B_3$ and $b_2$; for a convenient gauge--fixing of
the PST--symmetries the latter one amounts to
 \beq
 \label{BI}
 h_{ij}=-2{\delta{\cal L}\over \delta\widetilde h_{ij}}\equiv V_{ij} .
 \eeq

To write the action above one had to introduce a Chern--kernel
$K_4$, whose definition required two additional structures: normal
coordinates and an extension of the $SO(5)$--connection.
Eventually one has to make sure that the effective action does not
depend on the particular choice of these additional structures.
For a different choice of these structures we get a different
Chern--kernel $K^\prime_4$ which still satisfies
$dK^\prime_4=J_5=dK_4$. This means that $K_4$ and $K_4^\prime$
differ by an exact form. The requirement of independence of $H_4$
of the additional structures leads to the (finite) transformations
 \bea
  K_4^{\prime}&=&K_4 +dQ_3\label{prima}\\
  B_3^\prime &=& B_3 -\gbar Q_3\label{seconda},
  \eea
which imply
 \beq
  f_7^\prime =f_7+8K_4Q_3+4Q_3dQ_3+dQ_6, \quad
  K_4^\prime K_4^\prime={1\over 4}\,d\,f_7^\prime \label{quarta},
  \eeq
where
 $$
Q_3^{(0)} = 0 =Q_6^{(0)}.
 $$
These last identities follow from the fact that the behaviour of
$K_4$ near the 5--brane is universal, i.e. independent of the
choice of normal coordinates and of the extension of the
$SO(5)$--connection, and they ensure that $B_3^{(0)}$ and $\hb_3$
are independent of these structures, too.

The formula for $L_{11}$ reported above is completely fixed by the
requirement of invariance under \eref{prima}--\eref{quarta}.

\section{IIA $NS5$--brane  effective action\label{section:10D} }

The bosonic target space fields for IIA supergravity are a
three-form $A_3$, a two-form $A_2$, a one-form $A_1$, the dilaton
$\Phi $ and the metric $g_{\mu\nu}$, while on the bosonic
$NS5$--brane there live the coordinates $x^\mu (\sigma )$, a
two-form $a_2(\sigma )$ and a scalar $a_0 (\sigma )$.

In order to discuss the effective action for the system as a first
step one has to understand how the direct/back--coupling mechanism
works, i.e. one has to find  the coupled Bianchi identities and
equations of motions for the fields and to introduce potentials.

\subsection{The Chern--kernel}

The elementary $NS\, 5$--brane in ten dimensions, with charge $g$,
is described by a current $J_4$ that is the ten--dimensional
analog of $J_5$, and the normal bundle is now an $SO(4)$ one. The
basic Bianchi identity for the curvatures is the analog of eq.
\eref{eq:dH_4bis}:
 \begin{equation}
    dH_3 =g J_4 ,
    \label{eq:H_3bi1}
 \end{equation}
and again one has to solve the problem of defining a potential
$A_2$, or equivalently of finding an appropriate Chern--kernel.
The basic difference w.r.t. the eleven--dimensional case is that
for $J_4$ there exists no {\it invariant} three--form $K_3$ such
that $dK_3=J_4$. This difference is due to the fact that $J_5$ is
an odd current while $J_4$ is an even one and that the Euler
characteristic of an odd bundle is zero while the one of an even
bundle is non vanishing.

To find an appropriate form $K_3$ as in the previous case we
introduce a set of ten--dimensional normal coordinates $(\sigma
^i, u^r)$, $r=1,\dots,4$, such that the 5--brane stays at
$\vec{u} =0$. $J_4$ reads then
 \begin{equation}
    J_4 = \frac{1}{4!} \,\ve^{r_1\cdots r_4}du^{r_1}\cdots du^{r_4}
\delta^4(u),
    \label{eq:J_4_expression}
 \end{equation}
whatever is the particular set of normal coordinates used. We also 
pick up an arbitrary target--space extension $W^{rs}(\sigma,u)$ of
the normal bundle $SO(4)$--connection $W^{rs}(\sigma )$, with the
same requirements as in eq. \eref{eq:connection_extension}, and we
call the corresponding extended $SO(4)$--curvature $T=dW+WW$. In
terms of these extended objects and of ``hatted" coordinates
$\uh^r =u^r/\sqrt{u^2}$ one can define the following target--space
three--forms:
 \begin{eqnarray}
\chi_3 &=&\frac{1}{8(2\pi )^2 }\,\ve^{r_1\dots r_4}\left(
W^{r_1r_2}dW^{r_3r_4}+ \frac{2}{3}W^{r_1r_2}\left(
WW\right)^{r_3r_4} \right)  \label{eq:chi_3}
\\\omega_3 &=&-\frac{1}{2(2\pi )^2 }\,\ve^{r_1\dots
r_4}\uh^{r_1}D\uh^{r_2} \left( T^{r_3r_4}+
\frac{2}{3}D\uh^{r_3}D\uh^{r_4} \right).
\label{eq:omega_3}
 \end{eqnarray}
The form $\chi_3$ is a Chern--Simons form for an extended
Euler--characteristic, $d\chi_3=\chi_4$,
$$
\chi_4=\frac{1}{8(2\pi )^2 }\,\ve^{r_1\dots r_4}
T^{r_1r_2}T^{r_3r_4},
$$
and its pullback on the 5--brane is regular and coincides of
course with  the Euler Chern--Simons form on the brane. The
$SO(4)$--invariant three--form $\omega_3$ instead represents the
Chern--kernel associated to $J_4$ \cite{HarveyLawson}, that by
definition satisfies
 \begin{equation}
    d\omega_3 = J_4-\chi_4,
    \label{eq:d_omega_3}
 \end{equation}
as can be verified explicitly. The $J_4$--contribution in this
formula comes entirely from the ``pure Coulomb--form" $C_3$,
 \begin{equation}
C_3 = -\frac{1}{3(2\pi )^2}\,\ve^{r_1\dots r_4}\uh^{r_1}d\uh^{r_2}
d\uh^{r_3} d\uh^{r_4},\quad dC_3=J_4.
    \label{eq:C_3}
 \end{equation} 
As well as $K_4$, the form $\omega_3$ exhibits an invariant
singular behaviour near the 5--brane, although its pullback does
not exist. Eq. \eref{eq:d_omega_3} provides in particular a
realization of the Thom isomorphism \cite{BottTu}, i.e. the
cohomological equivalence of $J_4$ and $\chi_4$, the latter having
as pullback on $M_6$ indeed the Euler--form.

Thanks to \eref{eq:d_omega_3} we can choose for the three--form $K_3$
the combination
\begin{equation}
    K_3 = \omega_3+\chi_3 , \quad dK_3=J_4.
    \label{eq:K_3}
 \end{equation}
As in the eleven--dimensional case, the singular part of $K_3$
($\omega_3$) is invariant under $SO(4)$--transformations but now
$K_3$ has also a regular contribution ($\chi_3$) which transforms
under $SO(4)$ as
 \beq
 \label{transk3}
 \delta K_3 =d\chi_2.
 \eeq
We  can now introduce a potential two--form via
 \begin{equation}
    H_3 =dA_2 +gK_3,
    \label{eq:H_3bi_solution}
 \end{equation}
and we require $A_2$ to be regular on the brane. It is important
to realize that the introduction of a potential according to the
above equation does not represent the most general solution of
\eref{eq:H_3bi1}: it is the most general solution subject
to the boundary conditions represented by the universal singular
behaviour near the 5--brane exhibited by $K_3$; this is the physical
information we add. Notice also that $H_3$ does not admit pullback
on $M_6$, but in the Chern--kernel approach there is no need to
define this pullback, what is needed is the separation of $H_3$
into a regular part, $dA_2+g\chi_3$, and a singular one,
$g\omega_3$.

$SO(4)$--invariance of $H_3$ implies the anomalous transformation
law
 \begin{equation}
    \delta A_2 = -g\chi_2,
    \label{eq:deltaA_2bis}
 \end{equation}
which realizes in particular the pullback transformation law
\eref{trans} anticipated in the introduction.

In this case as well as in the previous one we have introduced 
additional structures, normal
coordinates and an extended $SO(4)$--connection. Under change of
these we have $K_3\rightarrow K_3^\prime$, $dK_3^\prime=J_4=dK_3$.
This means that $K_3$ and $K_3^\prime$ differ by a closed form,
and invariance of $H_3$ leads to the transformations
 \bea
    K_3^\prime &=& K_3 + dQ_2 \label{eq:deltaK_3}\\
    A_2^\prime &=& A_2 - g Q_2  \label{eq:deltaA_2}\\
    Q_2^{(0)}& =& 0. \label{eq:Q_2}
 \eea
An explicit expression for $Q_2$ as well as a proof of the last equation
are reported in Appendix A. This equation 
asserts that $Q_2$ is a regular target--space form, whose pullback
on the brane is well defined and equal to zero; this is again a
consequence of the fact that the Chern--kernel has a universal
singular behaviour near the 5--brane: apart from being
$SO(4)$--invariant this behaviour is independent of the choice of
normal coordinates and of the extension of the connection. As a
consequence $A_2^{(0)}$ is not only regular but also independent
of the additional structures. The transformations 
\eref{eq:deltaK_3} and \eref{eq:deltaA_2} should not be confused with
those under $SO(4)$--rotations \eref{transk3} and \eref{eq:deltaA_2bis}
of the normal bundle; in particular $\chi_2^{(0)}\neq 0$, opposite to  
$Q_2^{(0)}=0$.

The currents $J_5$ and $J_4$ exhaust all even and odd dimensional
currents. For a general $J_n$ one can introduce a form $K_{n-1}$
as
$$
J_n=dK_{n-1},\quad K_{n-1}=\omega_{n-1}+\chi_{n-1},
 \quad d\omega_{n-1}=J_n-\chi_n,
$$
where $\omega_{n-1}$ is an invariant Chern--kernel, and the Euler
characteristic of an odd bundle is zero by definition.

\subsection{Bianchi identities}

Now we are ready to discuss the remaining Bianchi identities. For
IIA {\it pure} supergravity they and the resulting potentials are
given by
 \begin{equation}
    \begin{array}{lcr}
    \left\{ \begin{array}{l}
                dH_1 =0 \\
                dR_2 =0 \\
                dH_3 =0  \\
                dR_4 = H_3 R_2  \\
            \end{array} \right.
                & \rightarrow  &
                 \left\{ \begin{array}{l}
                 H_1 =d\Phi \\
                 R_2 = dA_1 \\
                  H_3 = dA_2 \\
                  R_4 = dA_3 + H_3 A_1.\\
                  \end{array}
                  \right.
    \end{array}\label{pure}
 \end{equation}
When the $NS\, 5$--brane is present, the Bianchi identity for
$H_3$ gets modified and in order to keep the field algebra closed
w.r.t. differentiation the Bianchi identity for $R_4$ also 
receives a contribution of order $g$. The result is
 \begin{eqnarray}
    dH_1 &=&0 \label{eq:H_1bi} \\
    dR_2 &=&0 \label{eq:R_2bi} \\
    dH_3 &=&gJ_4 \label{eq:H_3bi2} \\
    dR_4 &=& H_3 R_2 -gh_1J_4 \label{eq:R_4bi} \\
    dh_1 &=& R_2^{(0)} \label{eq:h_1bi} \\
    dh_3 &=& \left(R_4 -H_3h_1\right)^{(0)} \label{eq:h_3bi10D} ,
 \end{eqnarray}
where we added the Bianchi identities for the brane potentials
$a_0$ and $a_2$.

Eqs. \eref{eq:H_3bi2}, \eref{eq:R_4bi} imply that the pullbacks of
$H_3$ and $R_4$ on the brane are ill--defined. However, in the
Bianchi identity \eref{eq:h_3bi10D} only the sum $R_4-H_3h_1$,
which is a closed form, is
required to have pullback, and we will see in a moment that this
combination is indeed regular. A part from this we should also mention
that the product $h_1H_3$ does not really define a target space
form since $h_1$ is only a field on the brane; these formal problems
are solved below.

Eqs. \eref{eq:H_1bi}, \eref{eq:R_2bi} and \eref{eq:h_1bi} can be
easily solved defining $H_1 =d\Phi $, $R_2 = dA_1$, $h_1 =d a_0
+A_1^{(0)}$, while the solution of eq. \eref{eq:R_4bi} requires
some caution. One would be led to write formally $R_4
=d\widetilde{A}_3 +H_3A_1-g a_0 J_4$, but this choice is not a
good one since $\widetilde{A}_3$ would not admit pullback. This can be
seen considering the transformation law $\delta A_1=d\Gamma$,
$\delta a_0=-\Gamma^{(0)}$
which requires $\delta \widetilde{A}_3=-H_3\Gamma$, and this does
not admit pullback. On the other hand, the spurious $\delta$--like
singularities in the term $a_0J_4$  should be cancelled by the
singularities present in $\widetilde{A}_3$.

A regular potential $A_3$ can be introduced considering the
alternative formal solution of eq. \eref{eq:R_4bi} $R_4
=dA_3+H_3h_1$, which does not exhibit any $\delta$--like
singularity. But this solution has a different problem: $h_1$ is
only a field living on the brane, and not a target--space one, so
the product $H_3 h_1$ is not well defined. Luckily there is a
solution for this problem: promote $h_1$ to a target--space form
$\widehat{h}_1$ by choosing an arbitrary extension
$\widehat{a}_0(x)$ of $a_0(\sigma )$ such that
 \begin{equation}
    \begin{array}{lr}
    \widehat{a}_0(\sigma ,0)=a_0(\sigma ), &
    \lim_{|u|\rightarrow\infty }\widehat{a}_0(\sigma ,u)=0,
    \end{array}
    \label{eq:a_0extension}
 \end{equation}
and by defining
$$
\widehat{h}_1(x) = d\widehat{a}_0(x)+A_1(x),
$$
with the properties
$$
\widehat{h}_1^{(0)}=h_1, \quad d\widehat h_1=R_2.
$$
This allows to
make use of the product $H_3\widehat h_1$ as a well defined target space form.
However, for consistency
the resulting $R_4$ should be independent of the choice of the
extension $\widehat a_0(x)$. Under a change of extension
$\widehat a_0$ we have indeed
 \begin{equation}
    \begin{array}{lr}
    \widehat a_0^{\,\prime} (x)= \widehat a_0 (x) +\Lambda (x),
    & \Lambda^{(0)}=0 ,
    \end{array}
    \label{eq:delta_aN1}
 \end{equation}
where the pullback of $\Lambda$ is zero due to eq.
\eref{eq:a_0extension}. This leads to
 \begin{equation}
    \left( H_3 \widehat h_1 \right)^\prime =H_3 \widehat h_1
     + H_3 d\Lambda =
    H_3 \widehat h_1 +d\left( H_3 \Lambda\right),
 \end{equation}
and $R_4$ remains unchanged if one chooses
\begin{equation}
     A_3^\prime = A_3 -H_3 \Lambda,
    \label{eq:delta_aN2}
\end{equation}
which has the remarkable property of leaving the pullback of $A_3$
invariant \footnote{The part of $H_3$ which does not admit pullback
is $g\omega_3$; the ``singularity" of $\omega_3$ is due to the fact that
as one approaches the 5--brane $u^r\rightarrow 0$,
$\omega_3^{(0)}$ while remaining {\it finite} would depend on the
direction along which $u^r$ goes to zero. Thus also $H_3$ remains finite,
and $H_3 \Lambda$ goes to zero as $u^r\rightarrow 0$ thanks to
\eref{eq:delta_aN1}.},
thanks to eq. \eref{eq:delta_aN1}. This is consistent
with the requirement that $A_3^{(0)}$ is regular and independent
of the chosen extension. We stress that having potentials whose pullbacks
are regular is an important part of our construction.

According to our definition of $\widehat h_1$ the r.h.s. of eq.
\eref{eq:h_3bi10D} has now to be replaced by $(R_4-H_3\widehat
h_1)^{(0)} =dA_3^{(0)}$, which is regular and closed; the solution
of this Bianchi identity amounts then simply to
$h_3=da_2+A_3^{(0)}$. We can now collect the solutions of our
Bianchi identities in terms of potentials,
 \begin{eqnarray}
    H_1 &=&d\Phi \\
    R_2 &=& dA_1 \\
    H_3 &=& dA_2 +gK_3 \label{sol}\\
    R_4 &=& dA_3 + H_3 \widehat h_1 \label{eq:A_3} \\
    h_1 &=& da_0 + A_1^{(0)}\\
    h_3 &=& da_2 + A_3^{(0)}.
 \end{eqnarray}
>From the definition of $h_3$ we deduce again the need of a
three--form potential with a regular pullback.

\subsection{Equations of motion}

The equations of motion for pure IIA supergravity are
\begin{eqnarray}
   d \left(e^{-2\Phi }* H_1 \right) &=&
 \frac{1}{4} R_4 * R_4 + \frac{3}{4} R_2 * R_2
 -\frac{1}{2} H_3 \left( e^{-2\Phi }* H_3 \right)\\
    d * R_2 &=& - H_3 * R_4  \\
    d \left( e^{-2\Phi } * H_3 \right) &=& \frac{1}{2}R_4 R_4
 -R_2 * R_4 \\d * R_4 &=& H_3 R_4,
\end{eqnarray}
where consistency requires the right hand sides of these equations to be
closed forms, as implied by the Bianchi identities \eref{pure}. In the
presence of a 5--brane, where the Bianchi identities change to
\eref{eq:H_1bi}--\eref{eq:h_3bi10D}, we have to change the equations
of motion as well to keep the r.h.s. closed. As in the case of the Bianchi
identities the new terms have to be supported on the 5--brane, i.e. to
be proportional to $J_4$. Our proposals for the new equations of $A_1$,
$A_2$ and $A_3$ are
 \begin{eqnarray}
    d * R_2 &=& - H_3 * R_4 + g J_4 * {\cal H}_1
    \label{first}\\
    d \left( e^{-2\Phi } * H_3 \right) &=& \frac{1}{2}R_4 R_4-R_2 * R_4
        + g h_1 h_3 J_4 +
    \frac{2\pi G}{g}\left(X_8 +\frac{1}{24}\chi_4 J_4\right)
    \label{eq:H_3eom} \\
    d * R_4 &=& H_3 R_4 - g h_3 J_4\label{last},
 \end{eqnarray}
while for the fields on the brane, $a_2$ and $a_0$, we propose
 \bea
h_{ij}&=&-2\,{\delta {\cal L}\over \delta\widetilde h_{ij}}
\label{a2}\\
d*{\cal H}_1&=&(*R_4-H_3\widehat h_3)^{(0)}\label{a0}.
 \eea
Before justifying the various new terms in these equations we
present the Bianchi identity and the equation of motion for $A_3$,
\eref{eq:R_4bi} and \eref{last}, in a different but equivalent
way. Notice first that the r.h.s. of \eref{last} is now closed,
because $d(H_3R_4)= d(H_3dA_3)=gJ_4dA_3=gJ_4dh_3$. Setting
$R_6=*R_4$ we can use this equation to introduce a dual potential
$A_5$. The dynamics of $A_3$ can then be represented equivalently
through the duality invariant system
 \bea
 R_4&=&dA_3+H_3\widehat h_1\label{a}\\
 R_6&=&dA_5+H_3\widehat h_3\label{b}\\
  R_6&=&*R_4\label{c},
 \eea
where we have extended the brane field $a_2$ to a target space form
$\widehat a_2$, in
exactly the same way as $a_0$ above, and we defined the target space
three--form
$$
\widehat h_3\equiv d\widehat a_2+A_3.
$$
Substituting \eref{c} in \eref{b} and
applying the differential one gets indeed back \eref{last}.

In analogy to $A_3$ we require also the dual potential $A_5$ to be
regular near the 5--brane, i.e. $A_5^{(0)}$ to be well--defined.
The self--consistency of this requirement proves in the same way
as in the case of $A_3$.  Its basic consequences, due to \eref{b} and
\eref{c}, are that
 \beq
 \label{suppl}
 d(*R_4-H_3\widehat h_3)=0, \quad
(*R_4-H_3\widehat h_3)^{(0)}\quad {\rm well}\,{\rm defined}.
 \eeq
One should notice that, although the statement that $A_3$ and $A_5$
admit pullback is duality symmetric, in the sense that the
two potentials are considered as equivalent, there is an intrinsic
asymmetry between $A_3$ and $A_5$ in $IIA$--supergravity, because there
is no formulation in terms
of $A_5$ only \footnote{There is clearly a lagrangian formulation for $IIA$
which involves {\it both} potentials, a la PST, in analogy to the $IIB$--case
\cite{Dallagata}.}. Our formulation is in terms of $A_3$ only, so
the requirement that $A_3$ admits pullback is of kinematical nature,
while the requirement that also $A_5$ admits pullback is a dynamical
constraint, because it asserts that only those solutions of the
$A_3$--equation of motion (eq. \eref{last}) are allowed, for which the
combination $*R_4-H_3\widehat h_3$ admits pullback. In conclusion,
the consistency of our system of equations of motion needs as
supplementary condition the second assertion in \eref{suppl}.

Keeping this in mind
we consider the equation for $A_1$, \eref{first}. We added on
its r.h.s. a term involving (the six--dimensional Hodge dual of) a
one--form on the brane, ${\cal H}_1$. Due to \eref{suppl} we have
\footnote{The
following computation has to be performed with some caution since a
straightforward application of Leibnitz's rule,
$d(H_3*R_4)=H_3d*R_4+dH_3*R_4=-gH_3h_3J_4+gJ_4*R_4$, leads to a
meaningless result because the pullback of $H_3$ does not exist.}
 $$
d(H_3 * R_4)= d(H_3(*R_4-H_3\widehat
h_3))=gJ_4(*R_4-H_3\widehat h_3)^{(0)},
 $$
from which it is clear that the term $gJ_4*{\cal H}_1$ is needed to make
the r.h.s. of \eref{first}
a closed form, thanks to the $a_0$--equation \eref{a0}. This last equation
is well--defined, again thanks to \eref{suppl}, and
it represents the $a_0$--equation of motion since at the linearized level
we expect
$$
{\cal H}_1=h_1+\cdots.
$$
The precise relation between ${\cal H}_1$ and $h_1$ will emerge
below when we write the action.

The $A_2$--equation \eref{eq:H_3eom} contains a term supported on
$M_6$, $gh_1h_3J_4$, which is needed to keep its r.h.s. a closed
form, as can be seen using the other Bianchi identities and
equations of motion. Also here, due to the presence of singularities,
one is not allowed to use Leibnitz's rule: the computation
$d({1\over 2}R_4R_4)=R_4dR_4=R_4(H_3R_2-gh_1J_4)$ makes no sense because
$R_4$ does not admit pullback. Rather one has first to evaluate $R_4R_4$
using the definition of $R_4$ in \eref{eq:A_3}, which gives
$R_4R_4=dA_3dA_3+2dA_3H_3\widehat h_1$, and then
$$
d\left({1\over 2}R_4R_4\right)=dA_3(H_3R_2-gh_1J_4)
=R_4H_3R_2-gdh_3h_1J_4;
$$
the rest of the computation is straightforward.  The terms proportional
to the ten--dimensional Newton's constant $G$ are closed by
themselves and are required for anomaly cancellation.

The equation of motion for the chiral two--form on the brane, eq.
\eref{a2}, is known from previous work \cite{BandosNurmaSorokin}.
It is governed by formally the same Born--Infeld lagrangian ${\cal
L}(\widetilde h)$ as the one of the $M5$--brane in \eref{BIL}. The
differences are that here we have $h_3=da_2+A_3^{(0)}$ and that
this time all six--dimensional indices in ${\cal
L}(\widetilde{h})$, in $\widetilde h_{ij}$ and in $h_{ij}$ are contracted
with the effective metric
 \begin{eqnarray}
    g_{eff,ij} &=& e^{-\frac{2}{3}\Phi }\left[ g_{ij}-e^{2\Phi }h_i h_j
    \right] \\
    g_{eff}^{ij}&=& e^{+\frac{2}{3}\Phi }\left[ g^{ij}+
    \frac{e^{2\Phi }h^i h^j }{1-e^{2\Phi }h^2 }\right],
    \label{eq:eff_metric}
 \end{eqnarray}
where $h_i$ are the components of the 5--brane one--form
$h_1=da_0+A_1^{(0)}$, and $h^2=h_ih_jg^{ij}$.
 A part from this we have again the
definitions $h_{ij}=v^k h_{ijk}$, $\widetilde{h}_{ij}=v^k \left(*
h\right)_{ijk}$ etc. as in the eleven--dimensional case.

These equations  of motion fix the classical action modulo terms
that are independent of $A_1$, $A_2$, $A_3$ and $a_2$, while the
equations of motion for $g_{\mu\nu}$ and $\Phi$, which are rather
complicated, are obtained varying the resulting action. This
action should fix also the dependence of ${\cal H}_1$ on $h_1$
and the other fields.

\subsection{Effective action}

The resulting effective action $\Gamma $ splits in a classical
action and a quantum part
$\Gamma_q $
 \begin{equation}
    \Gamma =\frac{1}{G}\left( S_{kin}+ S_{wz} \right) +\Gamma_q,
    \label{eq:10d_action}
 \end{equation}
where $\Gamma_q$ carries the anomaly associated to \eref{anomaly},
and the action $S_{kin}+ S_{wz}$ should give rise to the equations
of motion  of the previous section. Once the system of Bianchi
identities and equations of motion is consistent the
reconstruction of the action becomes a purely technical point.
Apart from invariance requirements under standard gauge transformations
for the potentials (as in pure supergravity) in the case at hand
one has to keep in mind that, due to our definition of $H_3$ in terms
of $K_3$, the action has also to be invariant under changes of $K_3$,
i.e. under the transformations \eref{eq:deltaK_3}--\eref{eq:Q_2}.
We give now the action and explain then how it produces our equations
of motion. The kinetic part is given by
 \bea
  S_{kin} &=& -\frac{1}{2}\int_{M_{10} } \left[
  R_4 * R_4 +  R_2 * R_2 +  e^{-2\Phi}\left(H_3 * H_3+8 H_1 * H_1\right)
  \right]\nonumber\\
  \phantom{S_{kin}}&& +\int_{M_{10}}d^{10}x \sqrt{g}\,R \,e^{-2\Phi }
  -g \int_{M_6}d^6 \sigma \sqrt{g_{eff} }\left({\cal L}(\widetilde{h})
  +\frac{1}{4}
  \widetilde{h}^{ij}h_{ij} \right)\label{kin10},
 \eea
and the Wess--Zumino term reads
 \begin{equation}
S_{wz} = -\frac{1}{2}\int_{M_{10} } A_3 dA_3 H_3-
\frac{g}{2}\int_{M_6} da_2 A_3 +
    \frac{2\pi G}{g} \left( \int_{M_{10} } H_3 X_7+ \frac{1 }{24}
    \int_{M_6} A_2^{(0)} \chi_4\right).
    \label{eq:WZ}
\end{equation}
In the last term of the kinetic action all indices are contracted with the
effective metric given above, but apart from this the dependence of
$S_{kin}+S_{wz}$ on the field $a_2$ is entirely fixed by the
PST--symmetries. After a convenient gauge--fixing of these
symmetries the $a_2$--equation of motion following
from this action is indeed \eref{a2}, see \cite{PST}.

In deriving the $A_3$--equation, the unique non trivial point is the
evaluation of the variation of the last term in $S_{kin}$ under a generic
variation of this field:
$$
\delta \int_{M_6}d^6 \sigma \sqrt{g_{eff} }\left({\cal
L}(\widetilde{h}) +\frac{1}{4} \widetilde{h}^{ij}h_{ij}\right)
=\int_{M_6}\left({1\over 2}\,h_3+v(h_2-V_2)\right)\delta A_3^{(0)}=
{1\over 2}\int_{M_6}h_3\delta A_3^{(0)},
$$
where $V_{ij}\equiv-2{\delta {\cal L}\over \widetilde h_{ij}}$,
and we used the (gauge--fixed) $a_2$--equation \eref{a2}. The variation
of the remaining terms is standard, and the resulting equation of motion
is \eref{eq:H_3eom}.

The derivation of the $A_2$--equation is straightforward since this field
does not appear in the last term of $S_{kin}$.

To derive the equations for $A_1$ and $a_0$, which show up
in the last term of $S_{kin}$ only in the combination $h_1$, we define
the 5--brane field
$$
{\cal H}_i\equiv-{1\over \sqrt{g}}
{\delta
\int_{M_6}d^6 \sigma \sqrt{g_{eff} }\left({\cal L}(\widetilde{h})
+\frac{1}{4} \widetilde{h}^{ij}h_{ij}\right)
\over \delta h^i}.
$$
This means that under a generic variation of $h_1$ we have
\beq
\label{generic}
\delta
\int_{M_6}d^6 \sigma \sqrt{g_{eff} }\left({\cal
L}(\widetilde{h}) +\frac{1}{4} \widetilde{h}^{ij}h_{ij}\right)
=-\int_{M_6}d^6 \sigma\sqrt{g}\,\delta h^i\, {\cal H}_i
=\int_{M_6}\delta h_1* {\cal H}_1,
\eeq
where the $*$-operation refers to the induced metric $g_{ij}$.
For a generic variation of $A_1$ this computation gives
$$
\int_{M_6}\delta h_1* {\cal H}_1=
\int_{M_6}\delta A_1^{(0)}* {\cal H}_1=\int_{M_{10}}
\delta A_1J_4* {\cal H}_1,
$$
which leads eventually to \eref{first}. For a generic variation of
$a_0$ instead we have
 \beq
 \label{vara0}
\int_{M_6}\delta h_1* {\cal H}_1=\int_{M_6}\delta a_0\,d(* {\cal H}_1).
 \eeq
To this one has to add the variation of the first term in $S_{kin}$,
which depends on the extended field $\widehat a_0$. Under a variation of
this field we have
 \bea
\delta\left(-{1\over 2}\int_{M_{10}}R_4*R_4\right)
&=& \int_{M_{10}} d\delta \widehat a_0\,H_3*R_4=
\int_{M_{10}}d\delta \widehat a_0\, H_3\left(*R_4-H_3 \widehat
h_3\right)\nonumber\\
&=& g\int_{M_6}\left(*R_4-H_3\widehat h_3\right)^{(0)}\delta a_0,
\nonumber
 \eea
where we used again \eref{suppl}. Despite the presence of the
extended field $\widehat a_0$ in the definition of $R_4$,
this variation is supported entirely on
$M_6$ and depends only on $\delta a_0$; this is clearly a consequence
of the independence of $R_4$ of the chosen extension. Adding this
variation to \eref{vara0}, multiplied by $-g$, one obtains the
equation for $a_0$. Finally, from the above definition of ${\cal H}_1$
it is easy to see that at first order in the fields it reproduces
$h_1$.

A further important point is that the classical action is
invariant under \eref{eq:deltaK_3}--\eref{eq:Q_2} because $A_2$
appears either in the combination $H_3=dA_2+gK_3$, or through its
pullback $A^{(0)}_2$ in  the anomaly cancelling term $\int
A_2^{(0)}\chi_4$, and both are invariant. The action is clearly
also invariant under standard gauge transformations of the
potentials.

Under $SO(1,9)$-- and $SO(4)$--rotations the Wess--Zumino action
transforms as (see \eref{eq:deltaA_2bis})
 \begin{equation}
    \delta \left({1\over G}S_{wz}\right)= - 2\pi \int_{M_6}
     \left(X_6 +\frac{1}{24}\chi_4
    \chi_2 \right),
 \end{equation}
which cancels against $\delta \Gamma_q$.

It is worthwhile  to notice that  to obtain the correct equation
of motion for $A_2$ it would have been sufficient to introduce the
$SO(1,9)$--invariant term $\int dA_2 X_7=\int A_2 X_8 $ in
$S_{wz}$, instead of $\int H_3 X_7$, but this term
alone would spoil the invariance
\eref{eq:deltaK_3}--\eref{eq:Q_2}. This forces the introduction of
the term $g\int K_3 X_7$, that is not invariant under $SO(1,9)$ and
cancels the target--space anomaly: invariance requirements and
anomaly cancellation are closely related.

\subsection{Dirac--branes}

Since the $NS5$--brane is electromagnetically dual to the $NS$--string
one can ask if the 5--brane dynamics can be treated in a way analogous
to the magnetic monopole in four dimensions which is, in turn, dual to
the electric charge. A necessary ingredient for a lagrangian
description of a magnetic monopole is the Dirac--string, i.e.
a two--dimensional surface whose boundary is the monopole worldline, an
ingredient which finally should result unobservable. This setup can be
generalized \cite{LechnerMarchetti} to a generic $p$--brane, where the
Dirac--string becomes a Dirac--(p+1)--brane. In this subsection
we illustrate briefly the corresponding setup for the $NS5$--brane, and
explain where {\it it eventually fails} in this case, due to the particular
dynamics of the $NS5$--brane.

With this respect the Chern--kernel
approach can be considered as equivalent to a Dirac--brane approach
{\it plus} a canonical superselection rule on the allowed curvatures $H_3$.

A Dirac--brane associated to the $NS5$--brane is a seven--manifold
$M_7$ whose boundary is $M_6$, $\pa M_7 =M_6$. Calling $C_3$ the
Poincar\`e--dual of $M_7$ this means
$$
dC_3 =J_4.
$$
Then one can give an alternative solution of equation \eref{eq:H_3bi2},
in terms of a potential $B_2$,
 \begin{equation}
 \label{sol1}
    H_3 =dB_2 +gC_3.
 \end{equation}
Up to here this solution and the previous one, $H_3=dA_2+gK_3$, are
equivalent because they correspond to a redefinition of the two--form
potential. Both solutions parametrize the most general solution of
$dH_3=gJ_4$: take a particular solution, $gK_3$ or $gC_3$, and add the
most general solution of the associated homogeneous equation.
However, taking the solution \eref{sol} in terms of $A_2$ we imposed
implicitly the ``canonical superselection rule" that $A_2$ is regular near
the 5--brane or, equivalently, that the singularities of $H_3$
near the 5--brane are the ones of the Chern--kernel. On the other hand,
the solution \eref{sol1} is completely general, and allows for example
also a configuration of the kind $H_3=gC_3$, which can clearly not be
realized in the Chern--kernel approach. In principle one could use also
\eref{sol1} and impose the constraint that $H_3$ behaves in the vicinity
of the 5--brane as $K_3$; this would amount to a very complicated constraint
on $B_2$, while it is extremely simple when imposed on $A_2$.

The question remains if one can do also without imposing any constraint
on $H_3$, using the most general solution \eref{sol1} in terms of a
Dirac--brane.
The basic problem one encounters in this case is that
$B_2$ can not admit pullback
on $M_6$. To see this we observe that the Dirac--brane is not unique;
under a change of Dirac--brane $M_7$ goes into
a new $M_7^\prime$ whose boundary is still $M_6$, so that
$M_7^\prime - M_7$ is boundaryless and there exists a 7--brane
$M_8$ such that $\pa M_8 = M_7^\prime - M_7$. In terms of the
corresponding Poincar\`e--duals we have $C_3^\prime-C_3=dC_2$,
and to have a Dirac--brane independent $H_3$ we must require
\begin{equation}
    B_2^\prime = B_2 - g C_2,
    \label{eq:delta_wt_A_2}
\end{equation}
which is analogous to \eref{eq:deltaA_2}. This time, however,
$C_2$ is a $\delta$--function on a manifold $M_8$ which contains as
submanifold $M_6$ and this implies that $C_2^{(0)}$ does not exist.
This means that a ``by--hand--requirement", that $B_2$ admits pullback on
$M_6$, is inconsistent. This would still not imply an inconsistency at the
level of equations of motion \footnote{There is, however, a problem related
with the transformation \eref{eq:delta_aN2}, since now the singularities of
$H_3$ could be even of the $\delta$--type and in this case the pullback of
$H_3\Lambda$ is not defined.}, but it would at the level of the action. 
Indeed, in the Wess--Zumino action ${1\over G}S_{wz}$ there is a term,
the normal--anomaly--cancelling term ${\pi\over12g}
\int_{M_6}A_2^{(0)}\chi_4$, in
which the  two--form does not appear in the combination $H_3$. This
should now be replaced by ${\pi\over12g}
\int_{M_6} B_2^{(0)}\chi_4$. But this term
has two problems: first, $B_2^{(0)}$ is not defined and second, $B_2$
depends on the Dirac--brane according to \eref{eq:delta_wt_A_2}.
To cope with the second problem -- Dirac--brane--dependence -- one
could replace this term by the formal expression
$$
    {\pi\over12g}\int_{M_6}H_3^{(0)}\chi_3,
$$
where $H_3^{(0)}=\left(dB_2+gC_3\right)^{(0)}$, since it gives rise to
the same equation of motion for the two--form. This would transform under
$SO(4)$, again formally, as
\begin{equation}
    \delta\left({\pi\over12g}\int_{M_6}
       H_3^{(0)}\chi_3\right)
=-{\pi\over12}\int_{M_6}J_4^{(0)}\chi_2.
     \label{eq:formal_change}
\end{equation}
But the pullbacks $H_3^{(0)}$, $C_3^{(0)}$  as well as
$J_4^{(0)}$ are ill--defined,
and one is back to the situation described in the introduction, where one
must invoke cohomological representatives.

In conclusion, the physical content of the Chern--kernel approach
is represented by the universal prescription for the singular behaviour
of the invariant curvature near the brane (superselection rule); with
this prescription one can
write a consistent set of Bianchi identities and equations of motion, and
an action which cancels the gravitational anomalies. The Dirac--brane
approach, instead, furnishes a framework which does not specify the
allowed singularities for the invariant curvature; it introduces, moreover,
intermediate unphysical $\delta$--like singularities along the
Dirac--brane. As a consequence it does not allow to write a
well--defined action.

In the next section we will perform a Kaluza--Klein dimensional
reduction of the system $M5$--brane + dynamical $D=11$  supergravity
of section \ref{section:11D}, down to ten dimensions. The result is the
system $NS5$--brane + dynamical $IIA$ supergravity, and we expect
to obtain our Chern--kernel formulation of the theory, because we
are starting from a theory written, in turn, in the Chern--kernel
formalism.

\section{Dimensional reduction \label{sec:dimensional_reduction}}

We perform in this section the dimensional reduction of the
$D=11$  $M5$--brane--action of section 2, down
to ten dimensions, compactifying say the coordinate $x^{10}$ on a circle of
radius $R$. The main motivation is clearly a consistency
check of the ten dimensional action constructed independently in the
previous section for the $NS5$--brane: the reduction process should
reproduce this action, modulo local terms.
The second motivation is related with the fact that in the present case the
reduction shows up a new (a priori problematic) feature w.r.t. the
reduction of pure sourceless $D=11$ supergravity, due to the presence of
the 5--brane.

We will indicate eleven--dimensional quantities with a bar and
ten--dimensional ones without bar; e.g. $x^{\muh}=(x^\mu,x^{10})$, where
$\mu=(0,\cdots,9)$. The relation between ten-- and eleven--dimensional
Newton's constants and magnetic charges is standard and reads
\beq
\label{const}
G= \frac{\Gbar }{2\pi R}, \quad  g= \frac{\gbar }{2\pi R}.
\eeq
Using the notations
of section 2 for $D=11$ fields and the ones of section 3 for $D=10$ fields,
we remember that the target space fields decompose as,
$g_{\muh\nuh}\rightarrow (g_{\mu\nu},A_1,\Phi)$ and
$B_3\rightarrow(A_3,A_2)$. The eleventh coordinate on the $M5$--brane
becomes the scalar field of the $NS5$--brane while the eleven--dimensional
chiral two--form is identified directly with the ten--dimensional one:
\bea\nonumber
x^{10}(\sigma)&=&a_0(\sigma)\\
b_2(\sigma)&=&a_2(\sigma).\nonumber
\eea
We recall first the standard decompositions of the eleven--dimensional
fields in the case of pure supergravity. The three--form decomposes as
 \beq
\label{standard}
 B_3=A_3+dx^{10}A_2,
 \eeq
while the metric is reduced according to
 \begin{eqnarray}
        E_{\muh}^{\overline{a} }&=&\left(
        \begin{array}{lr}
        e^{-\frac{1}{3}\Phi }E_\mu^{\, a} & e^{\frac{2}{3}\Phi }A_\mu
        \\0                                 & e^{\frac{2}{3}\Phi }
        \end{array}  \right)
        \rightarrow g_{\muh\nuh} =
        \left( \begin{array}{lr}
            e^{-\frac{2}{3}\Phi }
            \left(g_{\mu\nu}-e^{2\Phi }A_\mu A_\nu \right)
            & -e^{\frac{4}{3}\Phi }A_\mu \\
              -e^{\frac{4}{3}\Phi }A_\mu
              &-e^{\frac{4}{3}\Phi }
             \end{array} \right).
\label{eq:g_dec}
 \end{eqnarray}
In particular the eleventh component of the elfbein reads
$$
E^{10}\equiv dx^{\muh}E_{\muh}^{10}= e^{+\frac{2}{3}\Phi}(d\xten +A_1),
$$ where the gauge transformation of $A_1$ amounts to a
$D=11$ diffeomorphism of $x^{10}$, such that the combination
$dx^{10}+A_1$ is invariant.

The fields $A_1,A_2,A_3,g_{\mu\nu}$ and $\Phi$ are assumed to be periodic
in $x^{10}$ with period $2\pi R$. $D=10$, $IIA$ supergravity is obtained
if one keeps only
the zero modes of their Fourier expansion or, equivalently, if one assumes
all those fields to be independent of $x^{10}$. In this case
\eref{standard} leads
to the {\it invariant} decomposition (here we define $H_4\equiv dB_3$)
\beq
\label{free}
H_4=R_4+(dx^{10}+A_1)H_3,
\eeq
where the $x^{10}$--independent curvatures $R_4\equiv dA_3+dA_2A_1$ and
$H_3\equiv dA_2$ are the sourceless counterparts of the corresponding
ten dimensional curvatures of section 3, i.e. with $g=0=a_0$.

In the coupled case,
however, the same procedure can not be applied in a straightforward way
since we have $dH_4=\gbar J_5$
and the current $J_5$, even if in principle it admits a decomposition
like \eref{standard}, depends on {\it all} eleven coordinates $x^{\muh}$ in
a non trivial way and it is, in particular, intrinsically non periodic in
$x^{10}$. Since $dK_4=J_5$, the Chern--kernel inherits the same problematic
features from the current.

In this section we will show how one can
overcome these difficulties and get the relation \eref{free} also in
the coupled case -- in the limit $R\rightarrow 0$ -- where
$R_4$ and $H_3$ are replaced with their coupled expressions in section 3. The
validity of the decomposition \eref{free} is indeed the fundamental
ingredient of the reduction process, in the free as well as in the
coupled case.

\subsection{The reduced geometry in the presence of an $M5$--brane}

As we will see, in this case the parametrization of the reduced metric
\eref{eq:g_dec} can be kept unchanged, while the reduction of $B_3$
will be more complicated then \eref{standard}.

The first problem which arises in the presence of an $M5$--brane is that
it is inconsistent to restrict $x^{10}$ to a compact interval since
$x^{10}(\sigma)$ is identified with the 5--brane field $a_0(\sigma)$ which
is clearly unconstrained. A first suggestion to overcome this difficulty
would be
to consider as compact field the shifted variable $x^{10}-a_0(\sigma)$,
but this is not a target space field. As we will see below the right
choice for the compact variable is
\beq
x^{10}-\widehat a_0(x)\in \left[-\pi R,\pi R\right],
\eeq
where $\widehat a_0(x)$ is a ten--dimensional extension of
$x^{10}(\sigma)$,
which we identify  with the homonymous ten--dimensional field of
section 3. Notice that independence of a $D=11$ field of the
variable $x^{10}-\widehat a_0(x)$ is equivalent to independence of
$x^{10}$.

We can now address the reduction of the current $J_5$. If we indicate
with $J_4(x)$ the ten--dimensional current of the 5--brane, i.e. the
Poincar\`e--dual of the surface $x^\mu(\sigma)$ w.r.t. ten dimensions,
then we can rewrite the eleven--dimensional current as
$$
J_5=(dx^{10}-d\widehat a_0)\delta(x^{10}-\widehat a_0)J_4,
$$
which is independent on the chosen extension $\widehat a_0$ since $J_4$
projects it down to $x^{10}(\sigma)$. We are interested in a configuration
in which $D=11$ Sugra reduces to $D=10,IIA$ Sugra, corresponding to the
case in which all eleven--dimensional fields are taken to be independent
of $x^{10}$; the obstacle to this procedure introduced by $J_5$ is that,
despite the dependence of $J_4$ on only $x^\mu$,
$J_5$ depends also on $x^{10}$ trough the $\delta$--function. This
difficulty can
be solved by noting that in the limit $R\rightarrow 0$ we can replace
\beq
\label{replace}
\delta(x^{10}-\widehat a_0)\rightarrow {1\over 2\pi R},
\eeq
and
\beq
\label{ridotta}
J_5=d\left({x^{10}-\widehat a_0\over 2\pi R}\right) J_4,
\eeq
which has now the desired structure and allows a consistent
reduction of the equation $dH_4=\gbar J_5$.

The last problem one has to face is that in the eleven--dimensional
effective action the current $J_5$ enters also indirectly, through the
four--form
$K_4$, which is expressed in terms of normal coordinates. So we have
to determine the relation between the compact variable
$x^{10}-\widehat a_0$ and the $SO(5)$--normal coordinates $y^a$.
The choice of the eleventh coordinate as the compact one is of course
conventional and
corresponds more generally to a vector field $V^{\muh}\partial_{\muh}$;
in our case $V^{\muh}=(0,\cdots,0,1)$. On the 5--brane this vector field
decomposes in a tangent and in a normal component, meaning that the choice
of a compact variable identifies also a normal vector on the 5--brane
(the normal component of $V^{\muh}$), call it
$V^a(\sigma)$. Thus the normal group $SO(5)$ of the $M5$--brane
reduces to its  subgroup $SO(4)$ which leaves $V^a$ invariant.
Conventionally, through an $SO(5)$--rotation, we can choose
$V^a=(0,0,0,0,1)$ which means that we single out the fifth normal
coordinate $y^5=V^ay^a$ as an $SO(4)$--invariant one. We can then introduce
ten--dimensional $SO(4)$--normal coordinates $u^r$ $(r=1,\cdots,4)$
according to $y^a=(y^r\equiv u^r,y^5)$, and the
ten--dimensional $NS5$--brane current $J_4$ can then be expressed for
$R\rightarrow 0$ as in \eref{eq:J_4_expression},
$$
J_4 = \frac{1}{4!} \,\ve^{r_1\cdots r_4}du^{r_1}\cdots du^{r_4} \delta^4(u),
\quad J_5= dy^5\delta(y^5)J_4.
$$

We can now decompose $K_4$ using ten--dimensional normal
coordinates. The $SO(5)$--connection $A^{ab}$ is decomposed
according to $A^{ab}=(A^{rs}\equiv W^{rs}, A^{r5}\equiv L^r)$,
where $W^{rs}$ is the $SO(4)$--connection and $L^r$ is an
$SO(4)$--covariant vector. With these definitions the
Chern--kernel in equation \eref{eq:K_4a} allows the following
fundamental decomposition, whose derivation is lengthy but
straightforward:
 \beq K_4 = {1\over 2}\,\chi_4\,\yh^5
 -{1\over2}\,\omega_3\,d\yh^5 + dG_3,
 \label{eq:K_4_reduction}
 \eeq where
$\chi_4$ is the $SO(4)$ Euler--characteristic and $\omega_3$ is
the Chern--kernel associated to $J_4$, both defined in section 3.
$G_3$ is an $SO(4)$--{\it invariant} three--form whose explicit
expression is given in Appendix B; it involves $L^r$, $u^r$, $y^5$
and their $SO(4)$--covariant derivatives. We remember also that
$$
\hat y^5={y^5\over \sqrt{(y^5)^2+|\vec u|^2}}.
$$
As a check of the above decomposition we note that,
using the defining relation
for the $SO(4)$--Chern--kernel $d\omega_3=J_4-\chi_4$, the differential of
the r.h.s. of \eref{eq:K_4_reduction} reduces to
\beq
\label{check}
dK_4={1\over2}J_4d\hat y^5;
\eeq
but $J_4$ sets $\vec u=0$ and $\hat y^5$ gets replaced with the sign
function
$\epsilon(y^5)$; eventually $d\hat y^5\rightarrow
d\epsilon(y^5)=2dy^5\delta(y^5)$ and the
r.h.s. of \eref{check} amounts just to $J_5$.

Until now we have only
rewritten $K_4$ in the uncompactified theory. At the compactified
level $dK_4$ has to match moreover \eref{ridotta}; comparing this with
\eref{check} allows to relate eventually the compact
variable to the normal coordinates
\beq
\label{ident}
\hat y^5 ={1\over \pi R}\left(x^{10}-\widehat a_0\right).
\eeq
This relation is in particular consistent with the fact that $\hat y^5$
varies in the interval $[-1,1]$. It is understood that in the above
decomposition for $K_4$ the variable $\hat y^5$ is meant to be substituted
by this expression, even if we maintain as notation the symbol $\hat y^5$.
We recall that, as the replacement \eref{replace} is valid only for
$R\rightarrow 0$, also the above identification for $\hat y^5$ is valid
only in the same limit.
In terms of normal coordinates independence of $x^{10}$ amounts
then to independence of $y^5$, or equivalently of $\hat y^5$.

The structure of \eref{eq:K_4_reduction} allows now to
determine the reduction of  $B_3$ in the presence of
a 5--brane, which maintains \eref{free}. It is indeed easy to
see that the reduction
\beq
\label{eq:B_3_dec}
B_3=A_3-\pi R \hat y^5(dA_2+g\chi_3)-\gbar G_3,
\eeq
leads to
\beq
\label{final}
H_4=dB_3+\gbar K_4=R_4+(dx^{10}+A_1)H_3,
\eeq
where $H_3$ and $R_4$ are defined in \eref{sol} and \eref{eq:A_3}
respectively. The fields $A_3$ and $A_2$ defined in this way, as well
as the (extended) $SO(4)$--connection $W_{rs}$ showing up in $K_3$, are
eventually taken to be independent of $x^{10}$; this ensures
$x^{10}$--independence of $H_3$ and $R_4$, as in the sourceless
case. For pure Sugra,
$\widehat a_0=0=\gbar$, \eref{eq:B_3_dec} reduces to the standard
decomposition \eref{standard} modulo a gauge transformation.

For what
concerns invariances we observe that $B_3$ was invariant under
$SO(5)$ and so it is under $SO(4)$; since also $G_3$ is invariant under
$SO(4)$ the fields $A_3$ and $dA_2+g\chi_3$ have to be invariant under
this group separately, due to the presence of the factor $\pi R\hat y^5$.
This determines the expected anomalous transformation law for $A_2$,
$\delta A_2=-g\chi_2$, from the eleven--dimensional point of view.

The last ingredient we need is the reduction of the pullback
$B_3^{(0)}$. The second term in \eref{eq:B_3_dec} has vanishing pullback
due to \eref{ident}. From the explicit expression of $G_3$ in the appendix
\eref{eq:G_3_def}, one sees that it has actually a finite
pullback, but that it depends on the direction along which it is taken.
However, due to the fact that $G_3$ is multiplied by $\gbar=2\pi R g$,
its pullback gives no contribution for $R\rightarrow 0$
\footnote{Notice however that
one can not drop the term $\gbar G_3$ in \eref{eq:B_3_dec}, since the
action is cubic in $B_3$ and one could obtain finite contributions
from $G_3$, due to the relation \eref{quantum}}.
Thus in this limit we have
\beq
\label{limit}
B_3^{(0)}=A_3^{(0)}, \quad \hb_3=h_3=da_2+A_3^{(0)}.
\eeq

\subsection{Reduction of the action}

We are now able to perform the dimensional reduction of the
classical eleven--dimensional effective action of section two,
see equations \eref{kin} (kinetic terms) and \eref{eq:L_11} (Wess--Zumino).
The first should
reduce to $2\pi R$ times the ten--dimensional kinetic
terms in \eref{kin10}, and the integral $\int_{M_{11}}L_{11}$ should
reduce to $2\pi R$ times the ten--dimensional Wess--Zumino action
given in \eref{eq:WZ}. We perform the reduction
of the kinetic and Wess--Zumino terms separately.

\subsubsection{Kinetic terms}

The reduction of the Einstein--Hilbert term in \eref{kin} is standard
and requires only to insert the decomposition \eref{eq:g_dec}
$$
    \int_{M_{11}}d^{11}x \sqrt{\gbar}\,\overline{R} =
       2\pi R\left(\int_{M_{10}}d^{10}x e^{-2\Phi}\sqrt{g}R
    -\frac{1}{2}\int_{M_{10} }
    \left( R_2 * R_2 + 8 e^{-2\Phi }H_1 * H_1 \right) \right).
$$

The reduction of the Born--Infeld action
for the chiral two--form in \eref{kin}
has been anticipated above, see also \cite{BandosNurmaSorokin}.
We have $\hb_3=h_3$, and if one introduces the decomposition
\eref{eq:g_dec} in the eleven--dimensional induced metric one obtains
easily the ten--dimensional effective induced metric:
$\gbar_{ij}\equiv\partial_i x^{\muh}\partial_j x^{\nuh} g_{\muh\nuh}
=g_{eff,ij}$; the result is, by construction, the Born--Infeld lagrangian
of \eref{kin10}.

Due to the relation \eref{final}, which holds formally also
in sourceless supergravity, the reduction of the kinetic term for
the three--form is now standard, i.e. like in the case of
$D=11$ Sugra $\rightarrow$ $D=10,IIA$ Sugra:
$$
\int_{M_{11} }H_4 * H_4 =
    2\pi R \int_{M_{10}}\left(R_4 * R_4 +
     e^{-2\Phi}H_3 * H_3\right).
$$
The kinetic terms obtained this way, divided by $2\pi R$, coincide with
the ones of \eref{kin10}.

\subsubsection{Wess--Zumino term}

The reduction of the Wess--Zumino form $L_{11}$ is a little bit more
complicated. In principle one has to substitute the above expressions
for $B_3$ and $K_4$, \eref{eq:B_3_dec} and \eref{eq:K_4_reduction},
together with an analogous reduction for $f_7$ in $L_{11}$ and to
perform the computation. Actually one can take advantage from the invariance
of the eleven--dimensional Wess--Zumino action under
\eref{prima}--\eref{quarta} to simplify this computation. Its details are
given in the appendix, here we quote the result. Modulo a closed
form one gets
\beq
\label{l11}
{1\over2\pi R}\,L_{11}=-{1\over 2}A_3dA_3H_3 {d\hat y^5\over 2}
-{g\over 2}da_2A_3^{(0)}J_5
+{2\pi \Gbar\over \gbar}H_3X_7 {d\hat y^5\over 2}
+{\gbar^2\over 24}A_2\chi_4J_4 {d(\hat y^5)^3\over 2}.
\eeq
The integral over the eleventh coordinate is trivial,
$\int {d\hat y^5\over 2}=\int {d(\hat y^5)^3\over 2}=1$, and
${1\over 2\pi R}\int_{M_{11}}L_{11}$ coincides with the ten--dimensional
Wess--Zumino action in \eref{eq:WZ}, because thanks to the relations
\eref{quantum} and \eref{const} we have $\gbar^2=
{2\pi\Gbar\over \gbar}={2\pi G\over g}$.

\section{Concluding remarks}

The construction of the effective action for the interacting
system $NS5$--brane/$IIA$--supergravity presented in this paper is
based on a consistent solution of the magnetic equation
(Bianchi--identity) $dH_3=gJ_4$, in terms of a two--form $A_2$ and
a  Chern--kernel. In principle duality allows to circumvent this
problem by introducing a dual potential $A_6$ in which case this
Bianchi--identity would become an equation of motion. However,
there exists no formulation of $IIA$--supergravity in terms of
only $A_6$, therefore one has necessarily to solve the magnetic
equation. Form this point of view the 5--brane is really a {\it
dual} object. The magnetic equation allows
essentially for two classes of solutions, Dirac--branes or
Chern--kernels, the latter being in some sense a subclass of 
the former. As a
matter of fact you have to choose a Chern--kernel solution
whenever the {\it anomaly polynomial contains an Euler--form}, the
eleven--dimensional $M5$--brane being an exception. So the case
considered in this paper is a prototype for a rather general
situation: a dual or selfdual ($p$ or $D$)--brane with a normal
bundle Euler--form in the anomaly polynomial, see e.g.
\cite{CheungYin,LechnerMarchetti,Mourad,Scrucca}. The
Chern--kernel approach presented in this paper has general
validity and can be extended to all these cases: it leads to well
defined potentials and to a consistent anomaly cancelling
classical action. In particular, in the case of intersecting
$D$--branes the inflow mechanism is based at present on the {\it
cohomological} identification \cite{CheungYin}
 \beq
 \label{iden}
J_{M_1}J_{M_2}\sim J_{M_1 \cap M_2}\,\chi[N_{12}],
 \eeq
where $\chi[N_{12}]$ indicates the Euler characteristic of the
intersection of the normal bundles of the $D$--brane worldvolumes
$M_1$ and $M_2$, and $J_{M_i}$ denote the Poincar\'e duals of
$M_i$. From a {\it local} point of view (in the sense of pointwise) 
this formula has no
meaning since the product of currents $J_{M_1}J_{M_2}$ is not
defined, not even in the sense of distributions. Also in this case the 
Chern--kernel approach allows to overcome this
difficulty and to realize a local cancellation mechanism which
does not make use of the identification \eref{iden}, in analogy
with the $IIA$ $NS5$--brane where in our Chern--kernel
approach the analogous relation $J_4J_4\sim J_4\chi_4$ is never
enforced.  

 We have seen that even currents, like $J_4$, can be
written as the differential of an odd form, like $K_3$, that
transforms anomalously under the normal bundle group, while odd
currents like $J_5$ can be written as the differential of an
invariant form, $K_4$. In the first case the potential is
non--invariant while in the latter it is. In $D=10$ even currents
correspond to even brane worldvolumes where anomalies can
potentially appear explaining the appearance of anomalous forms
$K_{2n+1}$, realizing the inflow cancellation mechanism. Form this
point of view the appearance of an {\it invariant} Chern--kernel
$K_4$ in the case of the $M5$--brane seems rather strange. What
characterizes  eventually the exceptionality  of this case is that
eleven--dimensional supergravity has a Wess--Zumino term which is
cubic in the three--form potential and that the normal bundle
anomaly of the $M5$--brane does not factorize; moreover the Euler
characteristics of the normal bundle is zero. In the reduction
process from eleven to ten dimensions we have shown how those
different features are related in a non trivial way, in particular
the appearance of the non--invariant forms $K_3$ and $A_2$ from
the invariant ones $K_4$ and $B_3$.

\bigskip
\paragraph{Acknowledgements.} It is a pleasure
for the authors to thank P.A. Marchetti for very useful discussions.
This work is supported in part by the European Community's
Human Potential Programme under contract
HPRN-CT-2000-00131 Quantum Spacetime.

\bigskip
\section*{Appendix A: properties of the three--form $K_3$}

In this section we calculate the explicit expression of the form
$Q_2$ in equation \eref{eq:deltaK_3} and prove equation
\eref{eq:Q_2}, i.e. $Q_2^{(0)}=0$.

Starting from the definition of $K_3$, \eref{eq:chi_3},
\eref{eq:omega_3} and\eref{eq:K_3}, one can split $K_3$ in two
terms: the Coulomb--like form $C_3$, satisfying $dC_3 =J_4$, and
the differential of a two--form $\Phi_2$. An explicit calculation
yields
\begin{eqnarray}
    K_3 &=& C_3 + d\Phi_2\label{dec} \\
    \Phi_2(\uh,W) &=& -\frac{1}{4(2\pi )^2}\,\ve^{r_1\dots r_4} \uh^{r_1}
    \left( 2d\uh^{r_2} +(\uh W)^{r_2}
\right) W^{r_3r_4} \label{eq:Phi_2}\\
  C_3(\uh)&=&-\frac{1}{3(2\pi )^2}\,\ve^{r_1\dots r_4}\uh^{r_1}d\uh^{r_2}
d\uh^{r_3} d\uh^{r_4},
\end{eqnarray}
where, in particular, the form $C_3$ is independent of $W$.

We remember the anomalous transformation law of $K_3$ under
$SO(4)$ rotations, implied by the presence of the Chern--Simons
form $\chi_3$ in $K_3$. Under $SO(4)$ we have
\begin{equation}
    \begin{array}{lr}
    \uh^{\prime r}(\Lambda ) =\Lambda^{rs}\uh^s ,& W^\prime (\Lambda )
    = \Lambda W \Lambda^T - \Lambda d\Lambda^T ,
    \end{array}
    \label{eq:Wprime}
\end{equation}
where $\Lambda$ is an $SO(4)$--matrix,  and $K_3$ transforms as
\begin{equation}
    K_3(\uh^\prime ,W^\prime )= K_3(\uh ,W) +d\chi_2
(W,\Lambda),
    \label{eq:K_3rotation}
\end{equation}
where $d\chi_2 (W,\Lambda )$ is proportional to
$d\left[tr(\Lambda^{-1}d\Lambda W) \right] + \frac{1}{3} tr
\left(\Lambda^{-1}d\Lambda\right)^3$.

As stated in the text the form $K_3$ changes under a change of
normal coordinates and under a change of the extension of the
$SO(4)$--connection. In both cases we have $K^\prime-K_3=dQ_2$.
We consider first a change of normal coordinates. This amounts to a
transformation $u\rightarrow u^\prime (\sigma ,u)$ such that, (see
\cite{LechnerMarchettiTonin})
 \beq
u^{\prime r}(\sigma ,0) = 0,\quad  \left.\frac{\pa u^{\prime r}}
{\pa u^s}\right|_{u=0}= \delta^{rs}.
\label{norm}
 \eeq
Since  $\uh^{\prime r}\uh^{\prime r}=1$, there exists a matrix
$\Lambda(\sigma,u)\in SO(4)$ such that
 \beq
 \label{transform}
\uh^{\prime r} =\Lambda^{rs}\uh^s,
 \eeq
and \eref{norm} implies that for $\vec u\rightarrow 0$ we have the
crucial property
 \beq
    \Lambda^{rs} \sim\delta^{rs} + o(\vec u)f^{rs}(\uh)
    \Rightarrow\Lambda^{(0)}= 1.
\label{eq:first_transf}
 \eeq
For the new normal coordinate system we have, thanks to
\eref{eq:K_3rotation} and \eref{dec},
 \bea
    K_3^\prime = K_3(\uh^\prime,W) &=&
   K_3 (\uh, W^\prime (\Lambda^{-1}))
    +d\chi_2 (W^\prime(\Lambda^{-1}),\Lambda)\\
    &=& C_3(\uh)+d\Phi_2(\uh,W^\prime(\Lambda^{-1})
    +d\chi_2 (W^\prime (\Lambda^{-1}),\Lambda ).
 \eea
In the difference $K_3^\prime - K_3$ the Coulomb form cancels out
and we are left with 
\begin{equation}
    K_3^\prime - K_3 = d\left[ \Phi_2 (\uh, W^\prime (\Lambda^{-1}))
     - \Phi_2 (\uh, W) +
    \chi_2 (W^\prime (\Lambda^{-1}), \Lambda ) \right] = dQ_2 ,
\end{equation}
which gives an explicit formula for $Q_2$. Using \eref{eq:Phi_2},
and setting $\Delta W \equiv W^\prime (\Lambda^{-1}) - W$, we can
evaluate the difference
 \bea
&&\Phi_2(\uh,W^\prime(\Lambda^{-1}))-\Phi_2(\uh,W)\label{diff}\\
&&=-\frac{1}{4(2\pi)^2}\,\ve^{r_1\dots r_4} \uh^{r_1}
    \left(2d\uh^{r_2}\Delta W^{r_3r_4}+(\uh \Delta W)^{r_2}W^{r_3r_4}+
    \left(\uh W^\prime (\Lambda^{-1}) \right)^{r_2}\Delta W^{r_3r_4}
\right).\nonumber
  \eea
Thanks to \eref{eq:first_transf} we have now $\Delta W^{(0)}=0$,
and
$\chi_2^{(0)}(W^\prime(\Lambda^{-1}),\Lambda)=\chi_2(W^{(0)},1)=0$.
This implies also that $Q_2^{(0)}=0$.

Under a change of extension of $W$ instead we have
\begin{equation}
    \begin{array}{lr}
    W^\prime = W + \Delta W, & \Delta W^{(0)} = 0,
    \end{array}
    \label{eq:change_of_extension}
\end{equation}
since the value of $W$ on the brane is fixed. Accordingly  $K_3$
transforms as
\begin{equation}
    K_3^\prime = K_3(\uh , W^\prime ) = C_3(\uh ) +d\Phi_2
    (\uh ,W^\prime),
\end{equation}
so that
\begin{equation}
    K_3^\prime - K_3 = d\left[ \Phi_2 (\uh, W^\prime ) - \Phi_2 (\uh, W)
    \right] = dQ_2.
\end{equation}
Thus $Q_2$ coincides formally with \eref{diff} and $Q_2^{(0)}$
vanishes again, due to  $\Delta W^{(0)} = 0$.

\bigskip

\section*{Appendix B: the form $G_3$}

The decomposition of the invariant four--form $K_4$
\eref{eq:K_4_reduction} involves the $SO(4)$--invariant
three--form
\begin{equation}
    \begin{array}{ll}
    G_3 =
    \frac{1}{16(2\pi )^2 }\,\ve^{r_1\dots r_4} \uh^{r_1} &
    \left[ \frac{4}{3}\rho^2 \yh^5
    D\uh^{r_2}D\uh^{r_3}D\uh^{r_4}+4\rho\left( T^{r_2 r_3}+\rho^2
    D\uh^{r_2}D\uh^{r_3} \right)
    \right. L^{r_4} - \\ & \left.
    -4\rho^2 \yh^5 D\uh^{r_2}L^{r_3}L^{r_4} -\frac{4}{3}\rho^3
    L^{r_2}L^{r_3}L^{r_4}
    \right],
    \end{array}
    \label{eq:G_3_def}
\end{equation}
where
\begin{eqnarray}
    \uh^r &=& \frac{u^r}{|\vec{u}|}, \\
    \rho &=& \frac{|\vec{u}| }{\sqrt{|\vec{u} |^2+(y^5)^2}},
\end{eqnarray}
and $D\uh^r=d\uh^r+\uh^sW^{sr}$ indicates the covariant
derivative with respect to $SO(4)$. Since the ten--dimensional 5--brane is 
defined by $u^r=0$ we have $\rho J_4=0$, and therefore also $G_3J_4=0$ or, 
equivalently, the pullback of $G_3$ on the ten--dimensional 5--brane vanishes. 

\section*{Appendix C: reduction of $L_{11}$}

To reduce the eleven--dimensional Wess--Zumino action to ten
dimensions we have to evaluate the expression of $L_{11}$ in
\eref{eq:L_11} under the substitutions \eref{eq:B_3_dec} for $B_3$
and \eref{eq:K_4_reduction} for $K_4$, where eventually the forms
$A_3$, $A_2$, $\chi_3$ and $\omega_3$ are considered as
ten--dimensional fields, i.e. independent of the eleventh
coordinate.

Before performing the above substitutions it is convenient to eliminate
the form $G_3$ from $K_4$ and $B_3$, taking advantage from the invariance
of $L_{11}$ under the transformations \eref{prima}--\eref{quarta} and
choosing $Q_3=-G_3$. These transformations leave $L_{11}$ invariant if $Q_3$ 
has vanishing pullback on the eleven--dimensional 5--brane; in the reduced
geometry it is sufficient that it has vanishing pullback on the 
ten--dimensional 5--brane, as does $G_3$ (see appendix B). The new
objects simplify as
\bea
K_4^\prime &=& K_4 - d G_3 = \frac{1}{2}\chi_4\,
             \yh^5 -\frac{1}{2}\,\omega_3\,d\yh^5 \label{boh}\\
B_3^\prime &=& B_3 + \gbar G_3 = A_3  -\pi R \yh^5\left(dA_2+g\chi_3\right).
\eea
For what concerns the transformed seven--form $f^\prime_7$ it is sufficient
to determine it modulo a closed form $df_6$, since it would modify the
Wess--Zumino action by a term which is local on the 5--brane,
$\int_{M_{11}}df_6K_4=-\int_{M_6}f_6$ 
\footnote{We remind that the $D=11$ theory carries an $SO(5)$--anomaly
associated to $P_8(A)$. Under dimensional reduction we have 
$A^{ab}=(A^{rs}\equiv W^{rs}, A^{r5}\equiv L^r)$, and one has the 
decomposition $P_8(A)=\chi_4(W)\chi_4(W)+dl_7$, where $l_7$ is some
$SO(4)$--invariant form. For the Chern--Simons forms one has
$P_7=\chi_3\chi_4+l_7 +dl_6$, implying that the $SO(5)$--anomaly
reduces to the canonical $SO(4)$--anomaly $\int\chi_2\chi_4$,  
{\it modulo} a trivial cocycle. Our specific choice of $f_7^\prime$  
amounts to the subtraction of a local counterterm which restores the
canonical form of the anomaly, as seen from \eref{test}.}.
This means that it is sufficient to find a solution of the equation
$$ 
{1\over 4}df^\prime_7=K_4^\prime K_4^\prime  
=   {1\over 4}(\hat y^5)^2\chi_4\chi_4 
        +{1\over 2}\,\hat y^5 d\hat y^5 \omega_3\,\chi_4,
$$             
for  $f_7^\prime$. A convenient solution is 
\beq
\label{f7}
f_7^\prime=\left((\hat y^5)^2\chi_4-\omega_3\,d(\hat y^5)^2\right)\chi_3.
\eeq
Modulo a closed form we can therefore rewrite the eleven--form in
\eref{eq:L_11} by replacing $B_3\rightarrow B_3^\prime$, $K_4\rightarrow
K_4^\prime$, $f_7\rightarrow f_7^\prime$,
\begin{eqnarray}
  L_{11}&=&\frac{1}{6}B_3^\prime dB_3^\prime dB_3^\prime
 -\frac{\gbar}{2}\,d a_2 A_3^{(0)}J_5
+\frac{\gbar}{2}\,B_3^\prime dB_3^\prime K_4^\prime + \nonumber \\
&+& \frac{\gbar^2}{2}\,B_3^\prime K_4^\prime K_4^\prime 
   + \frac{\gbar^3}{24}K_4^\prime f_7^\prime +
   {2\pi\Gbar\over\gbar}\overline{X}_7 H_4. 
\label{l_11}
\end{eqnarray}
The substitution of \eref{boh}--\eref{f7} is now a mere exercise,
and the result can be divided into powers of $A_3$: the term 
$A_3dA_3dA_3$ vanishes of course because it is an eleven--form 
in $D=10$; the terms quadratic in $A_3$ lead easily to the first term
in \eref{l11}, while the linear terms drop out apart from the term
$da_2A_3^{(0)}J_5$. The interesting terms are the ones without $A_3$,
since they are related to anomalies and we evaluate them separately.
The first, third and fourth terms in \eref{l_11} produce a contribution
\beq
\label{uno}
{\gbar^2(\hat y^5)^3\over 48}\left(\gbar\,\omega_3\chi_4\chi_4   
-2\pi R\left(dA_2+g\chi_3\right)\chi_4J_4
\right),
\eeq
which is invariant under $SO(4)$, as are the three terms from which it 
comes. The fifth term instead, which carries the anomaly,
becomes
\beq
\label{due}
{\gbar^3\over 24}K_4^\prime f_7^\prime =
{\gbar^3(\hat y^5)^3\over 48}\left(\chi_3\chi_4J_4-
\omega_3\chi_4\chi_4\right).
\eeq 
We can check that it carries still the correct $SO(4)$--anomaly (the 
second term is invariant):
\beq
\label{test}
\delta\left({\gbar^3\over 24}\int_{M_{11}}K_4^\prime f_7^\prime\right)
=-{\gbar^3\over 24}\int_{M_{11}}\chi_2\chi_4J_4 {d(\hat y^5)^3\over 2}.
\eeq
This formula maintains its meaning both from the eleven-- and
ten--dimensional points of view. In $D=11$ $\hat y^5$ is given
by $\hat y^5={y^5\over \sqrt{(y^5)^2+|\vec u|^2}}$ which multiplied by 
$J_4$ reduces to the sign of $y^5$, hence $J_4{d(\hat y^5)^3\over 2}=
J_4dy^5\delta(y^5)=J_5$, and \eref{test} reduces to $-{\gbar^3\over 24}
\int_{M_6}\chi_2\chi_4$. From the ten--dimensional point of view we have to
take $\chi_4$ and $\chi_2$ to be independent of $\hat y^5$ and we
can integrate $\int_{-1}^1 {d(\hat y^5)^3\over 2}=1$, and one obtains
the same anomaly as before. 

Summing up \eref{uno} and \eref{due} only the $A_2$--term survives
and, apart from a closed form, one gets as anomaly cancelling term
$$
2\pi R\left({\gbar^2\over 24}A_2\chi_4J_4 {d(\hat y^5)^3\over 2}
\right),
$$ 
which amounts to the last term in \eref{l11}. 

The reduction of the last term in \eref{l_11}, which
carries the target space anomaly, is standard. 
We can decompose \footnote{This decomposition holds again modulo a closed
form, see the footnote on page 27.}
the $SO(1,10)$ Chern--Simons form as 
$\overline{X}_7=X_7+Z_7+ (dx^{10}+A_1)Z_6$, where $X_7$, $Z_7$ and
$Z_6$ are independent of $x^{10}$; $X_7$ is the $SO(1,9)$ Chern--Simons
form, and $Z_7$ and $Z_6$ are $SO(1,9)$--{\it invariant}. 
Recalling also the decomposition $H_4=R_4+(dx^{10}+A_1)H_3$ one gets
$$
\overline{X}_7H_4=\left(H_3X_7+ [H_3Z_7+ R_4Z_6]\right)dx^{10}.
$$
The term $H_3Z_7+R_4Z_6$ is local and invariant and corresponds to a non 
minimal contribution to the ten--dimensional effective action, while 
$H_3X_7dx^{10}= \pi R H_3X_7d\hat y^5$ produces the third term in 
\eref{l11}.


\vskip1truecm

\begin{thebibliography}{99}

\bibitem{Witten:fivebrane} E. Witten, J. Geom. Phys. {\bf 22}
(1997) 103.

\bibitem{BottTu} R. Bott and L. W. Tu, {\it Differential forms in
algebraic geometry}, Springer Verlag 1978.

\bibitem{CheungYin} Y. K. Cheung and Z. Yin, Nucl. Phys. {\bf B517}
(1998) 69.

\bibitem{BeckerAndBecker} K. Becker and M. Becker, Nucl. Phys. {\bf B577}
(2000) 156.

\bibitem{HarveyMinasianMoore} D. Freed, J.A. Harvey, R. Minasian and
G. Moore, Adv. Theor. Math. Phys. {\bf 2} (1998) 601.

\bibitem{LechnerMarchetti} K. Lechner and P.A. Marchetti, JHEP {\bf 01}
(2001) 003.

\bibitem{Gueven} R. Gueven, Phys. Lett. {\bf B276} (1992) 49.

\bibitem{LechnerTonin} K. Lechner and M. Tonin, Nucl. Phys. {\bf B475}
                        (1996) 545.

\bibitem{LechnerMarchettiTonin} K. Lechner, P.A. Marchetti and M. Tonin,
      Phys. Lett. {\bf B524} (2002) 199.

\bibitem{deRham} G. de Rham, {\it Differentiable manifolds, Forms,
Currents, Harmonic Forms}, Springer Verlag 1984.

\bibitem{HarveyLawson} F. R. Harvey and H. B. Lawson Jr.,
{\it A theory of characteristic currents associated with a
singular connection}, Asterisque 213 (1993), Soci\`et\`e
Math\`ematique de France.

\bibitem{morgan} F. Morgan, {\it Geometric Measure Theory}, Academic
             Press, San Diego, California (1988).

\bibitem{sigma_model} I. Bandos, K. Lechner, A. Nurmagambetov, P. Pasti,
D. Sorokin and M. Tonin, Phys. Rev. Lett. {\bf 78} (1997) 4332.

\bibitem{PST} P. Pasti, D. Sorokin and M. Tonin, Phys. Lett.
{\bf B398} (1997) 41.

\bibitem{Dallagata} G. Dall'Agata, K. Lechner and M. Tonin,
JHEP, {\bf 9807} (1998) 017.

\bibitem{BandosNurmaSorokin} I. Bandos, A. Nurmagambetov and
D. Sorokin, Nucl. Phys. {\bf B586} (2000) 115.

\bibitem{Mourad} P. Brax and J. Mourad, Phys. Lett. {\bf B416}
(1998) 295; J. Mourad, Nucl. Phys. {\bf B512} (1998) 199.

\bibitem{Scrucca} C. Scrucca and  M. Serone, Nucl. Phys. {\bf B556}
(1999) 197.


\end{thebibliography}
\end{document}